\documentclass[aps,prc,twocolumn,groupedaddress]{revtex4-2}

\usepackage{graphicx}
\usepackage{slashed}
\usepackage{color}
\usepackage{amsmath}
\usepackage{multirow}
\usepackage{xspace}
\usepackage{bm}

\usepackage{placeins} % for FloatBarrier
\usepackage[normalem]{ulem} % for sout

\newcommand{\nopieft}{$\slashed{\pi}$EFT\xspace}

\newcommand{\be}{\begin{equation}}
\newcommand{\ee}{\end{equation}}

\begin{document}
%=============================================================================
\title{Charged Particle Scattering in Renormalizable Pionless Effective Field Theory at Next-to-Leading Order: The $pd$, $dd$, and $p^3\mathrm{He}$ Case}
%\title{Power-counting renormalizable Pionless EFT up to next-to-leading order with non-perturbative Coulomb interaction: $\rm p-d$, $\rm d-d$, and $\rm p- ^3He$ scattering}
\author{Mat\'{u}\v{s} Rojik}
\email{mrojik@uni-mainz.de}
\affiliation{Nuclear Physics Institute of the Czech Academy of Sciences, \v{R}e\v{z} 25068, Czech Republic}
\affiliation{Faculty of Mathematics and Physics, Charles University, Prague 116 36, Czech Republic}

\author{Martin Sch\"{a}fer}
\email{m.schafer@ujf.cas.cz}
\affiliation{Nuclear Physics Institute of the Czech Academy of Sciences, \v{R}e\v{z} 25068, Czech Republic}

\author{Mirko Bagnarol}
\affiliation{The Racah Institute of Physics, The Hebrew University of Jerusalem, Jerusalem 9190401, Israel}

\author{Nir Barnea}
\affiliation{The Racah Institute of Physics, The Hebrew University of Jerusalem, Jerusalem 9190401, Israel}

\date{\today}

\begin{abstract}

We formulate a renormalizable pionless effective field theory (\(\slashed{\pi}\)EFT) with a non-perturbative treatment of the Coulomb interaction up to next-to-leading order (NLO) for few-nucleon systems. We extract scattering observables for charged clusters by employing two-, three-, and four-body contact interactions and using the stochastic variational method with a Coulomb-corrected harmonic oscillator trap. 
Our NLO results yield a \(pd\) spin-quartet scattering length and effective range of \(a_{pd}^{3/2} = 12.76(29)\,\mathrm{fm}\) and \(r_{pd}^{3/2} = 1.17(7)\,\mathrm{fm}\); for \(dd\) scattering in the spin-quintet channel, we find \(a_{dd}^{2} = 6.26(3)\,\mathrm{fm}\) and \(r_{dd}^{2} = 1.41(7)\,\mathrm{fm}\); and for \(p^3\mathrm{He}\) scattering, the spin-singlet and spin-triplet channels are characterized by \(a_{p^3\mathrm{He}}^0 = 11.26(4)\,\mathrm{fm}\), \(r_{p^3\mathrm{He}}^0 = 1.65(26)\,\mathrm{fm}\) and \(a_{p^3\mathrm{He}}^1 = 9.06(4)\,\mathrm{fm}\), \(r_{p^3\mathrm{He}}^1 = 1.36(25)\,\mathrm{fm}\), respectively. 
Our predictions exhibit mild cutoff dependence and agree well with existing experimental phase shift analyses and potential model calculations. This demonstrates the predictive power of \(\slashed{\pi}\)EFT for charged few-nucleon systems.

\end{abstract}

%\begin{keyword}
%Nuclei, Pionless Effective Field Theory, NLO, few-body scattering, Coulomb interaction
%\end{keyword}

\maketitle
%=====================
\section{Introduction}
%=====================
Effective field theories have caused a revolution in our understanding of nuclear interaction \cite{eft_review_2020}. They offer a framework consistent with the symmetries of quantum chromodynamics and simultaneously allow the description of processes at low energies characteristic for nuclear physics.  Based on the pioneering work of Weinberg \cite{WEINBERG1,WEINBERG2}, modern nuclear potentials used in \emph{ab initio} calculations of nuclear properties and reactions are constructed from chiral effective field theory ($\chi$EFT) \cite{machleidt2024review}, which includes contact interactions and explicit exchanges of pions. 

We focus on the simplest nuclear EFT, the pionless effective field theory (\nopieft) \cite{kolck1999}. Following the EFT paradigm, the theory is described by the most general Lagrangian involving only the nucleonic fields, which are the sole degrees of freedom. Since the breakdown scale of \nopieft is set by the pion mass $m_\pi$, this theory is particularly fit for describing few-nucleon systems and processes, where the typical momenta are much lower. In the past, it has been successfully applied to two-nucleon scattering and deuteron ($d$) properties \cite{nn_scattering_eft_1999,kong2000coulomb} and by using auxiliary dibaryon fields, the studies were extended to the three-body sector, the triton and helion, and $nd/pd$ scattering \cite{rupak2003quartet,vanasse2014he,konig2015proton,konig2016_3H_3He}. The three-nucleon bound states were also investigated using a potential formulation of \nopieft within the resonating group method \cite{kirscher_3He}. Recently, up to five-nucleon scattering was studied \cite{schafer2023few,bagnarol_letters_b} using renormalizable potentials derived from \nopieft at next-to-leading order. However, these calculations did not take the Coulomb interaction into account. 

The problem with \nopieft is that at sufficiently low momenta, Coulomb effects between protons dominate over the strong interaction. This means the power counting scheme must be adjusted to accommodate the new scale. This challenge was first tackled by Kong and Ravndal in $pp$ scattering \cite{kong1999proton,kong2000coulomb}. They showed that for typical nucleon momenta $Q <\alpha~m_N/2$, where $m_N$ is the nucleon mass {and $\alpha$ is the fine structure constant}, Coulomb needs to be treated non-perturbatively. {A s}imilar approach was followed by Vanasse \emph{et al.} in their study of $^3$He and $pd$ scattering \cite{vanasse2014he}. Alternatively, the Coulomb interaction can be treated as a perturbative effect, as was done in the study of $pd$ spin-quartet scattering by Rupak and Kong \cite{rupak2003quartet} and later extended to the $pd$ spin-doublet scattering and $^3$He bound state by K\"{o}nig and Hammer \cite{konig2011low}. Both approaches to the Coulomb interaction were compared in a calculation of the $pd$ spin-quartet scattering length by the same authors in Ref. \cite{konig_pd}, and the three-nucleon system in \nopieft was then revisited in Refs. \cite{konig2015proton,konig2016_3H_3He}. 

The studies mentioned in the previous paragraph are based on a Lagrangian formulation of \nopieft with auxiliary dibaryon fields. This complex technique has not been applied to nuclear systems involving more than three particles to date. The development of various techniques for solving the few-body Schr\"{o}dinger equation, such as the no-core shell model with resonating group method (NCSM/RGM) \cite{ncsm_rgm}, NCSM with continuum (NSCMC) \cite{ncsmc}, Faddeev-Yakubovsky equations \cite{fadeev_review}, or hyperspherical harmonics (HH) \cite{hyperspherical_review} has made it possible to obtain very accurate results of low-energy scattering properties of various projectiles off light nuclei. This enables the assessment of the accuracy of different potential models or properties of various EFTs. Despite this recent development, charged few-body scattering has not yet been explored thoroughly within the potential formulation of \nopieft. So far, only leading order $p^3$He scattering was calculated in Ref.~\cite{kirscher2013zero}. Regarding higher orders, Coulomb effects were considered in Refs .~\cite {lensky2016description,schiavilla2021}, where only bound-state few- and many-body nuclear properties were studied up to next-to-next-to-next-to-leading order of \nopieft. Since, in these works, the authors resummed higher orders in the Hamiltonian, the calculations were inconsistent with the renormalizable \nopieft power counting. 

In this work, we construct a renormalized \nopieft interaction up to next-to-leading order (NLO) employing local regulators. The Coulomb interaction is treated non-perturbatively, and only the leading order (LO) of the strong interaction is resummed; NLO effects are treated perturbatively. We build upon our previous studies done in Refs.~\cite {schafer2023few,bagnarol_letters_b} by additionally considering the Coulomb interaction. This allows us to extend the study to the scattering of charged nuclei. We present a comprehensive study of three-body ($pd$) {and} four-body ($p^3$He, $dd$) scattering within \nopieft. We use the stochastic variational method (SVM) \cite{suzuki2003stochastic} with a correlated Gaussian basis to solve the Schr\"{o}dinger equation together with a Coulomb-corrected harmonic oscillator trap technique \cite{guo2021coulomb} to obtain the Coulomb-subtracted phase shifts. Our EFT results are renormalization-group invariant, and we compare them to available experimental data and potential model calculations. 

This work is structured as follows - in Section~\ref{subsec:model} we provide the details on \nopieft and in Sec.~\ref{subsec:fit} we describe the process of fitting the low-energy constants (LECs) of the corresponding LO and NLO potential terms. We then discuss the methods used for obtaining the scattering phase shifts and solving the Schr\"{o}dinger equation in Sec.~\ref{subsec:method}. Our results are described in Sec. \ref{sec:results} and summarized in Sec. \ref{sec:conclusion}. The Appendix lists fitted values of \nopieft LO and NLO LECs used throughout this work.

%========================
\section{Theory}\label{sec:theory}
%========================
\subsection{Pionless EFT with non-perturbative Coulomb interaction up to next-to-leading order}\label{subsec:model}
In  \nopieft, the only degrees of freedom are the nucleonic fields themselves, and the corresponding effective Lagrangian density can be written as an expansion in contact interaction terms \cite{eft_review_2020}.
In the absence of the Coulomb interaction, the LO two-body potential obtained from such a Lagrangian can be written as
\begin{align}\label{lo}
    V^{(0)}_\text{2B}=\sum_{i<j}\left(C_0^{(0)}(\Lambda)\hat{\mathcal{P}}_{ij}^{(0,1)}+C_1^{(0)}(\Lambda)\hat{\mathcal{P}}_{ij}^{(1,0)}\right)g_\Lambda(r_{ij}),
\end{align}
where $\hat{\mathcal{P}}_{ij}^{(S,I)}$ is the projection operator on to the two-body $s$-wave channels with spin $S$ and isospin $I$. The $g_\Lambda(r_{ij})=\exp(-\Lambda^2r_{ij}^2/4)$ is the local Gaussian regulator which smears the contact interaction terms over the distance $\Lambda^{-1}$, and $r_{ij} = |{\bf r}_i - {\bf r}_j|$ stands for the relative distance between nucleons $i$ and $j$. Upon regularization, the low-energy constants $C_0^{(0)}$, $C_1^{(0)}$ gain specific dependence on the cutoff $\Lambda$. Physical observables must be independent of this cutoff; this is achieved by renormalization, i.e., by fitting the LECs to a selected set of low-energy experimental data.

To renormalize the $A>2$ nuclear systems at LO, the momentum-independent three-body force needs to be introduced into the three-body $s$-wave spin-doublet channel \cite{BEDAQUE2000357,PhysRevLett.82.463}. Here, we use its cyclic form 
\begin{equation}
    V_\text{3B}^{(0)}=\sum_{i<j<k}D_0^{(0)}(\Lambda)\sum_{cyc}\hat{\mathcal{P}}_{ijk}^{(1/2,1/2)}g_\Lambda(r_{ij})g_\Lambda(r_{ik}), 
\end{equation}
where $cyc$ stands for the cyclic sum over the $\{ijk\}$, $\{kij\}$, $\{jki\}$ triplets,  $\hat{\mathcal{P}}_{ijk}^{(1/2,1/2)}$ is a projection operator on to the three-body $(S,I)=(1/2,1/2)$ channel, and $D_0^{(0)}$ is the corresponding LEC. 

The NLO correction to the LO potential comprises two-body momentum-dependent (derivative) terms. Additionally, a contact four-body force must be introduced to renormalize the spatially-symmetric four-body channel \cite{4body_bosons, schafer2023few}. The NLO correction reads 
\begin{align}\label{potnlo}
    V^{(1)} =& \sum_{i<j}\left(
        C_0^{(1)}(\Lambda)~\hat{\mathcal{P}}_{ij}^{(0,1)}
       +C_2^{(1)}(\Lambda)~\hat{\mathcal{P}}_{ij}^{(1,0)}
       \right)g_\Lambda(r_{ij})
  +\nonumber\\
    +& \sum_{i<j}\left(
     C_1^{(1)}(\Lambda)~\hat{\mathcal{P}}_{ij}^{(0,1)}
    +C_3^{(1)}(\Lambda)~\hat{\mathcal{P}}_{ij}^{(1,0)}
    \right)\times
    \nonumber\\
    &~~\times \left(g_\Lambda(r_{ij})\overset{\rightarrow}{\nabla}^2
        +\overset{\leftarrow}{\nabla}^2g_\Lambda(r_{ij})\right)+
        \nonumber\\[5pt]
    +& \sum_{i<j<k} D_0^{(1)}(\Lambda)\sum_{cyc}\hat{\mathcal{P}}_{ijk}^{(1/2,1/2)}
    g_\Lambda(r_{ij})g_\Lambda(r_{ik})+
    \nonumber\\[5pt]
    +& 
    \sum_{i<j<k<l} E_0^{(1)}(\Lambda)
    ~\hat{\mathcal{P}}^{(0,0)}_{ijkl}~g_\Lambda(r_{ijkl}), %\label{eq:4b_nlo}
\end{align}
where the terms with $C_0^{(1)}$, $C_1^{(1)}$, and $D_0^{(1)}$ LECs stand for the NLO counterterms, $\hat{\mathcal{P}}^{(0,0)}_{ijkl}$ is the projection operator on to the spatially-symmetric four-body $(S,I)=(0,0)$ channel, and $r^2_{ijkl}=\sum_{a<b\in\{i,j,k,l\}}r_{ab}^2$ is the four-body hyperradius. 

While the LO potential $V^{(0)}=V^{(0)}_\text{2B}+V^{(0)}_\text{3B}$ is iterated through all orders via the Schr\"{o}dinger equation, the NLO correction $V^{(1)}$ is included within first-order perturbation theory. In such a way, momentum-dependent contact terms do not cause renormalization problems due to the Wigner bound \cite{Wigner1955}.  

The distinct feature of our work is a non-perturbative inclusion of the Coulomb interaction. Following the previous studies in Refs.~\cite{kong2000coulomb,vanasse2014he}, a new LEC must be introduced at LO to renormalize the $pp$ $s$-wave properly. In practice, we split the two-body spin-singlet, isospin-triplet $NN$ channel in Eq.~(\ref{lo}) into the $nn$, $np$, and $pp$ components. While the $C_0^{(0)}$ LEC enters only the $nn/pn$ channels, the $pp$ part of $V^{(0)}_\text{2B}$ is now described using the Coulomb interaction plus the additional momentum-independent $pp$ term
\begin{align}\label{2bc_lo}
    V^{(0)}_{pp}=\sum_{i<j}C_2^{(0)}(\Lambda)~\hat{\mathcal{P}}_{ij}^{(0,1;pp)}g_\Lambda(r_{ij}) + \frac{e^2}{r_{ij}}. 
\end{align}
Here, $\hat{\mathcal{P}}_{ij}^{(0,1;pp)}$ is the projection operator into the $(S,I)=(0,1)$ $pp$ channel. Equivalently to the LO case, we treat separately the $s$-wave $pp$ channel also at NLO     
\begin{align}\label{2bc_nlo}
    V_{pp}^{(1)}=&\sum_{i<j}C_4^{(1)}(\Lambda)~\hat{\mathcal{P}}_{ij}^{(0,1;pp)}g_\Lambda(r_{ij})+\nonumber\\
    +&\sum_{i<j}C_5^{(1)}(\Lambda)~\hat{\mathcal{P}}_{ij}^{(0,1;pp)}\left(g_\Lambda(r_{ij})\overset{\rightarrow}{\nabla}^2+\overset{\leftarrow}{\nabla}^2g_\Lambda(r_{ij})\right)
\end{align}
and introduce two new LECs $C_4^{(1)}$, $C_5^{(1)}$ into the $V^{(1)}$ correction.

%3b
The non-perturbative inclusion of the Coulomb interaction does not require a new three-body force at LO. However, an additional $ppn$ three-body force must be included at NLO to properly renormalize the theory \cite{vanasse2014he,konig2015proton}. Here, we add    
\begin{equation}\label{3bc_nlo}
    V_{ppn}^{(1)}=\sum_{i<j<k}D_1^{(1)}(\Lambda)\sum_{cyc}\hat{\mathcal{P}}_{ijk}^{(1/2,1/2;ppn)}g_\Lambda(r_{ij})g_\Lambda(r_{ik}),
\end{equation}
term to $V^{(1)}$, Eq.~(\ref{potnlo}). The $\hat{\mathcal{P}}_{ijk}^{(1/2,1/2;ppn)}$ denotes the projection operator on to the $ppn$ $(S,I)=(1/2,1/2)$ three-body channel. Upon inclusion of $V_{ppn}^{(1)}$, the $D_0^{(1)}$ term in Eq.~(\ref{potnlo}) enters in our work only for the $(S,I)=(1/2,1/2)$ $pnn$ three-body channel. 

%=========================================================================================
\subsection{Fitting the LECs}\label{subsec:fit}

In total, thirteen LECs must be fixed to low-energy data for a chosen range of cutoff values. At LO, $C_0^{(0)}$ is fitted to reproduce the experimental $nn$ spin-singlet scattering length $a^0_{nn}=-18.95$~fm \cite{gonzalez2006neutron, chen2008measurement} defined by the effective range expansion \cite{bethe1949theory}
\begin{equation}\label{eq:nocoulomb_ere}
    \kappa(\eta=0) \equiv k~\text{cotg}\left(\delta\right)=-\frac{1}{a}+\frac{1}{2}rk^2+\mathcal{O}(k^4),
\end{equation}
where $\delta$ is the $s$-wave phase shift, $k=\sqrt{2 \mu E}$ the relative momentum, and $r$ the effective range. Due to the lack of Coulomb interaction, the dimensionless Sommerfeld parameter $\eta=\mu e^2/k$, where $\mu$ is the reduced mass, is set to zero. The $C_1^{(0)}$ LECs is fixed to the deuteron binding energy $B({}^2\text{H})=2.2246$~MeV~\cite{vanderlauen1982}. The $C_2^{(0)}$ LEC is fixed to the experimental $pp$ spin-singlet scattering length $a^0_{pp}=-7.8063$ fm \cite{PhysRevC.38.15}, which is defined by the Coulomb-modified effective range expansion \cite{bethe1949theory}
\begin{align}\label{eq:coulomb_ere}
    \kappa(\eta) &\equiv \nonumber\\
    &C_0^2(\eta)~k~\text{cotg} (\delta)+2 k \eta~h(\eta)=-\frac{1}{a}+\frac{1}{2}r k^2+\mathcal{O}(k^4),
\end{align}
where $C_0^2(\eta)=\frac{2\pi\eta}{e^{2\pi\eta}-1}$, and $h(\eta)=\text{Re}[\Psi(1+i\eta)]-\ln(\eta)$ with $\Psi$ being the logarithmic derivative of the gamma function. 

The NLO correction introduces six additional two-body LECs. They are fitted to reproduce experimental effective ranges of the channels discussed $r^0_{nn}=2.75\,\text{fm}$ \cite{vslaus1989neutron}, and $r^0_{pp}=2.794\,\text{fm}$ \cite{PhysRevC.38.15} while keeping the corresponding scattering lengths intact. For the $np$ spin-triplet channel, we use the effective range value defined by the effective range expansion around the deuteron binding momentum above $r^1_{pn}=1.764\,\text{fm}$ \cite{nn_scattering_eft_1999}, while keeping $B({}^2\text{H})$ intact.  

The LO three-body LEC, $D_0^{(0)}$, is fitted to the $\rm ^3H$ binding energy $B(^3\text{H})=8.482$ MeV \cite{purcell2010energy}. At NLO, the three-body LEC $D_0^{(1)}$ enters only the $pnn$ channel, and it is fixed to keep $B(^3\text{H})$ intact at this order. The $D_1^{(1)}$ three-body LEC enters only the $ppn$ channel and it is fitted to $B(^3\text{He})=7.718$ MeV \cite{purcell2010energy}. Finally, the NLO four-body LEC $E_0^{(1)}$ is fixed to the $\rm ^4He$ binding energy, $B(^4\text{He})=28.296$ MeV \cite{tilley1992energy}. In our calculations, we keep the same proton and neutron masses $m_p=m_n=m_N$, use the nucleon mass parameter $(\hbar c)^2/m_N = 41.471073$~MeV.fm$^2$, {and the elementary charge squared 
$e^2 = \alpha \hbar c = 1.44$ MeV.fm.}

In the absence of the Coulomb interaction, we fit the $C^{(0)}_0$ using the variable phase method \cite{ouerdane2003variable} and $C^{(0)}_1$ using the Numerov algorithm. The $pp$ $C_2^{(0)}$ LEC is fixed by extracting scattering phase shifts at very low momenta from the calculated Coulomb-modified Jost function \cite{newton2013scattering}. The two-body NLO LECs are fitted using the distorted-wave Born approximation. 
We fix the $A>2$ LECs to experimental bound state energies using the Stochastic Variational Method (SVM) {\cite{suzuki2003stochastic}}, briefly outlined in the following Subsec.~\ref{subsec:method}. The LO three-body LEC is fitted by calculating $B(^3\text{H})$. The three- and four-body NLO LECs are fixed by calculating the first-order perturbative correction to the corresponding LO bound-state energies. We list all values of LECs used in this work in Tab.~\ref{tab:lo_lecs_2b},~\ref{tab:nlo_lecs_2b},~and~\ref{tab:nlo_lecs_3b_4b} in the Appendix.

\begin{figure*}
    \begin{tabular}[c]{cc}
            \includegraphics[width=0.5\linewidth]{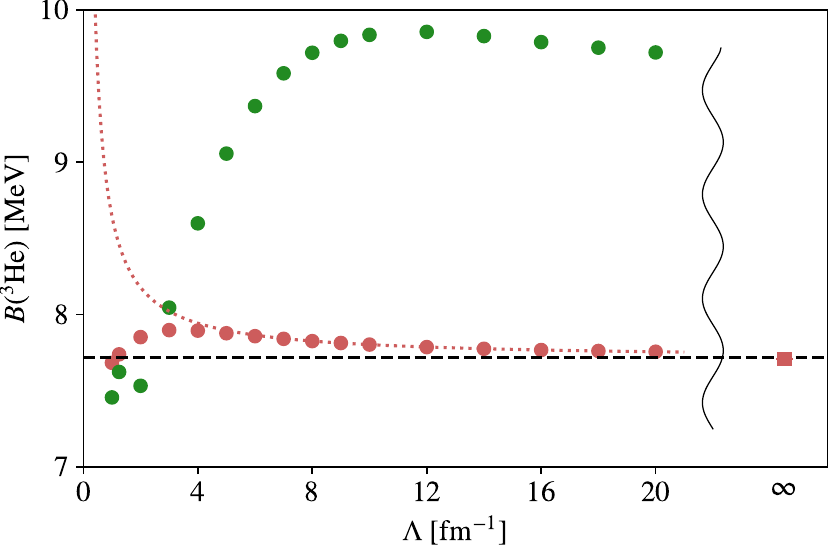}&
            \includegraphics[width=0.5\linewidth]{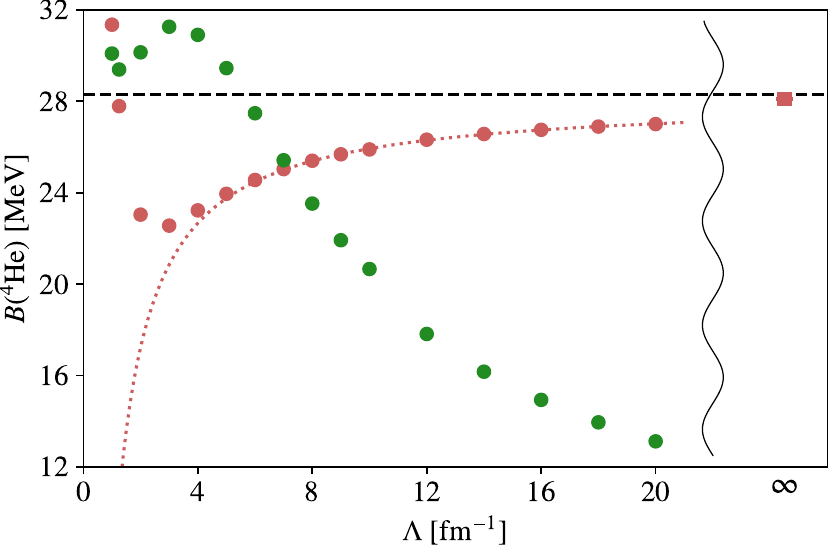}
    \end{tabular}
    \caption{The binding energies $B(^3\text{He})$ (left panel) and $B(^4\text{He})$ (right panel) as a function of the cutoff $\Lambda$. In both panels, the LO results are denoted by the red circles, and the red dotted lines show the fitted LO extrapolation function, Eq.~\eqref{extrapol}. The extrapolated LO values for $\Lambda\rightarrow\infty$ are denoted by the red squares. For $B(^3\text{He})$, the green circles show the results of our NLO calculations once the $pnn$ $D^{(1)}_0$ term is considered in place of the $ppn$ $D^{(1)}_1$ term. For $B(^4\text{He})$, the green circles denote calculations with all NLO correctly accounted for except the four-body force. Experimental values \cite{purcell2010energy,tilley1992energy} are represented by the black dashed lines.}
    \label{binding_energs}
\end{figure*}

%============================================================================================
\subsection{Few-body methods}\label{subsec:method}
We employ the harmonic oscillator (HO) trap method corrected for Coulomb long-range interactions to extract the elastic scattering phase shifts of charged $A>2$ systems. Two interacting clusters are trapped in the HO potential 
\begin{equation}
    V_\text{HO}=\frac{m_N}{2A}\omega^2\sum_{i<j}(\bm{r}_i-\bm{r}_j)^2, 
\end{equation}
where $\omega$ is the HO frequency. Provided the trap is sufficiently larger than the size of the clusters and the range of the short-ranged nuclear interaction, the $s$-wave phase shifts $\delta$ of two scattered clusters are given by \cite{guo2021coulomb}
\begin{align}\label{bush_c}
    -2\mu&~C_0^2(\eta)~k\text{cotg}\left(\delta\right)=\nonumber\\
    &\lim_{r,r'\to0}\left\{\text{Re}\left[G^{C,\infty}_{0}(r,r';E)\right]-G^{C,\omega}_{0}(r,r';E)\right\},
\end{align}
where $G^{C,\omega}_{0}$, $G_0^{C,\infty}$ are the Coulomb Green's functions of clusters in the HO trap and a free space, respectively. The Green's functions implicitly depend on the cluster charges $q_1$ and $q_2$. The $E$ is the bound-state energy with respect to the breakup threshold of the two clusters. For more details on the phase shift extraction and the numerical methods, we refer to our previous work {\cite{bagnarol_letters_b}}.

%NLO

To solve the bound-state few-body Schr\"{o}dinger equation, we expand the total wave function into a correlated Gaussian basis {\cite{Varga1998, Suzuki2008}}
\begin{equation}
    \Psi^{L=0}_{S M_S;I M_I}=\sum_i c_i \hat{\mathcal{A}}\left\{\exp\left(-\frac{1}{2}\bm{x}^T\bm{A}_i\bm{x}\right)\chi^i_{SM_S}\xi^i_{IM_I}\right\},
\end{equation}
where $\hat{\mathcal{A}}$ is the antisymetrization operator, $\bm{x}^T = (\bm{x}_1,\dots,\bm{x}_{N-1})$ is a vector of Jacobi coordinates, $\chi^i_{SM_S}$ and $\xi^i_{IM_I}$ are the spin and isospin part of the total wavefunction, respectively, and $\bm{A}_i$ is a symmetric positive definitive matrix which consists of $A(A-1)/2$ nonlinear basis state parameters. The optimal values of parameters are chosen by the Stochastic Variational Method \cite{suzuki2003stochastic}. The variational parameters $c_i$ are obtained by solving the generalized eigenvalue problem.

%==============================================================================
\section{Results}\label{sec:results}
%==============================================================================
We apply the above-outlined theory and methods to study the $s$-wave scattering of charged three-nucleon ($pd$) and four-nucleon ($dd$, $p^3$He) systems. All observables are calculated for a large range of cutoff values. Results in the $\Lambda \rightarrow \infty$ contact limit  are obtained by fitting our LO results with 
\begin{equation}\label{extrapol}
    f(\Lambda)={f_\infty} +\frac{c}{\Lambda} 
\end{equation}
and our NLO results using 
\begin{equation}\label{extrapolNLO}
    f(\Lambda)={f_\infty} +\frac{c}{\Lambda^2} 
\end{equation}
for $\Lambda \geq 4~\text{fm}^{-1}$. Here, $f(\Lambda)$ is the value of an observable at a finite cutoff $\Lambda$ and ${f_\infty}$ is the corresponding value extrapolated into the contact limit. 

The predominant source of uncertainty in our calculations is the theoretical error that arises from truncating the \nopieft Lagrangian expansion at the given order. These errors are assessed by studying the residual cutoff-dependence \cite{griesshammer2020consistency}. More specifically, for $\Lambda \geq 4~\text{fm}^{-1}$, we evaluate the spread $\Delta_\text{theor.}$ between the largest and smallest value of an observable $f(\Lambda)$ calculated in the studied cutoff range, including ${f_\infty}$. If not denoted otherwise, the numerical errors of $f(\Lambda)$, induced by the few-body methods, are negligible and not shown in our figures. However, they are considered in the extrapolation procedure. Our final result at the specific \nopieft order is then given as ${f_\infty (\Delta)}$, where the total uncertainty $\Delta = \sqrt{\Delta^2_\infty +\Delta^2_\text{theor.}}$ is given by the theoretical one and the extrapolation error, $\Delta_\infty$.

%--------------------------------------------------
\subsection{Binding energies of $^3$He and $^4$He}
%--------------------------------------------------

\begin{figure}
    \centering
    \includegraphics[width=\linewidth]{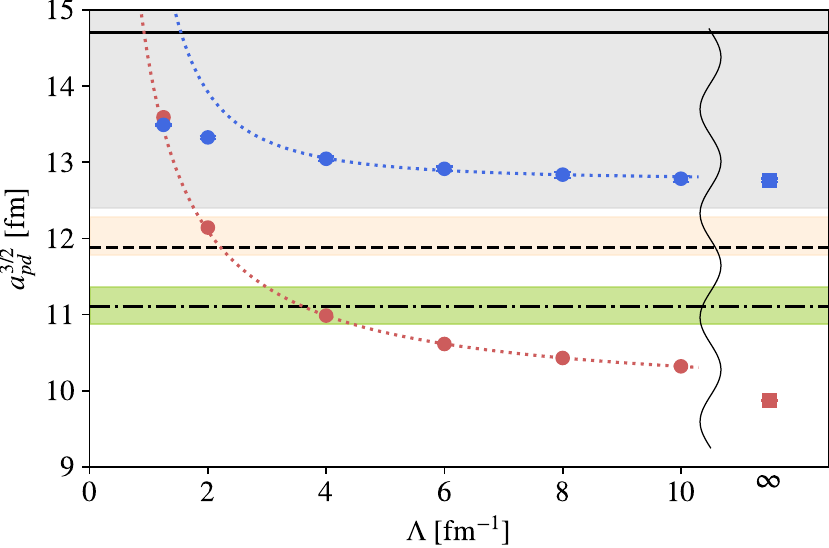}
    \caption{The $pd$ scattering length in the $S=3/2$ channel, $a^{3/2}_{pd}$, displayed as a function of the momentum cutoff $\Lambda$. The LO results are denoted by the red circles and the NLO by the blue circles. The dotted red (LO) and blue (NLO) lines denote the extrapolating functions Eqs.~\eqref{extrapol}~and~\eqref{extrapolNLO}, respectively. Corresponding extrapolated values at $\Lambda\rightarrow\infty$ are separated from the rest of the calculations by the wavy line and are denoted by squares. The solid black line and the grey band represent the most recent experimental 1999 datum and its error \cite{pd_scatt_length_black}, the dashed-dotted line and green band the 1983 datum \cite{huttel1983phase}, and the dashed line with the light orange band the 1974 datum \cite{arvieux1974phase}.}
    \label{pd_32_scatt_length}
\end{figure}

To test the adopted power counting scheme, i.e., the necessity for the $pp$ and $ppn$ Coulomb-related terms Eqs.~\eqref{2bc_lo}, \eqref{2bc_nlo}, and \eqref{3bc_nlo}, we first study the ground state binding energy of the helion and the alpha particle. The left panel of Fig.~\ref{binding_energs} shows our $B({}^3\text{He})$ results calculated for cutoffs $\Lambda\in[1,20]$ fm$^{-1}$. At LO, the results are obtained with the $pp$ term, Eq.~\eqref{2bc_lo}, which is iterated in the Hamiltonian. We observe that for $\Lambda \geq 4$~fm$^{-1}$ the binding energy converges. We obtain $B(^3\text{He})=7.7085(4)$~MeV by extrapolating the LO results to the contact limit, in excellent agreement with the experimental value of $ 7.718$~MeV. The error in the parentheses stands for $\Delta_\infty$ only.

At NLO, we consider all perturbative two-body corrections, as described in Subsec.~\ref{subsec:model}, but instead of the $V^{(1)}_{ppn}$ term, Eq.~\eqref{3bc_nlo}, we try to use the $pnn$ term with $D_0^{(1)}$ LEC, renormalized to the $B({}^3\text{H})$ ground state binding energy. Within this scheme, the corresponding NLO $B({}^3\text{He})$ energies (green points) deviate rapidly from the experimental datum with increasing momentum cutoff and, in the studied cutoff range, they show no apparent sign of the inverse powers of $\Lambda$ behavior. This observation is in agreement with the earlier study in Ref.~\cite{vanasse2014he}, where the ground state binding energy of $^3$He was calculated analytically using the dibaryon formalism. Here, the authors pointed out that the new $ppn$ term is necessary at NLO to cancel the logarithmic $\Lambda$ divergences induced by the nonperturbative inclusion of the Coulomb interaction.  

The right panel of Fig.~\ref{binding_energs} presents the calculated ground state binding energy of $^4$He as a function of $\Lambda$. Similarly as for the ${}^3$He case, the calculated LO $B(^4\text{He})$ starts to converge for $\Lambda \geq 4$~fm$^{-1}$ and our extrapolation into the contact limit yields $B(^4\text{He})=28.11(3)$~MeV, in good agreement with the experimental datum 28.296~MeV. The error in the parentheses gives an extrapolation error only. The inclusion of the Coulomb interaction and the $pp$ and $ppn$ contact terms does not affect the necessity of the four-body NLO force. This is demonstrated by the green points, which show the calculated $B(^4\text{He})$ energies with all NLO terms, except the four-body force, perturbatively accounted for. The cutoff dependence of the resulting energies is comparable to the previous non-Coulomb study in Ref.~\cite{schafer2023few}: for $\Lambda \geq 4~\text{fm}^{-1}$, the calculated NLO binding energy starts to decrease rapidly, showing no apparent sign of inverse powers of $\Lambda$ behavior, and eventually largely deviates from the LO result or the corresponding experimental datum.      
\begin{figure*}
    \begin{tabular}[c]{cc}
            \includegraphics[width=\columnwidth]{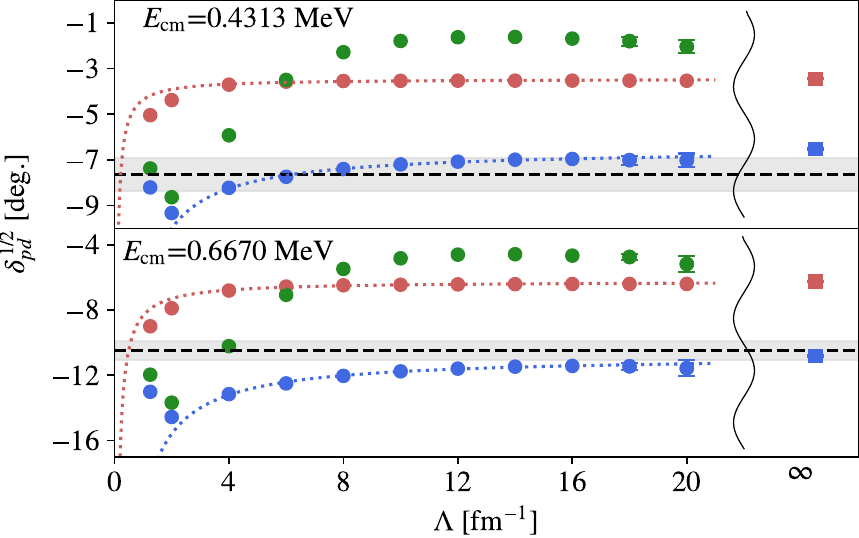}&
            \includegraphics[width=\columnwidth]{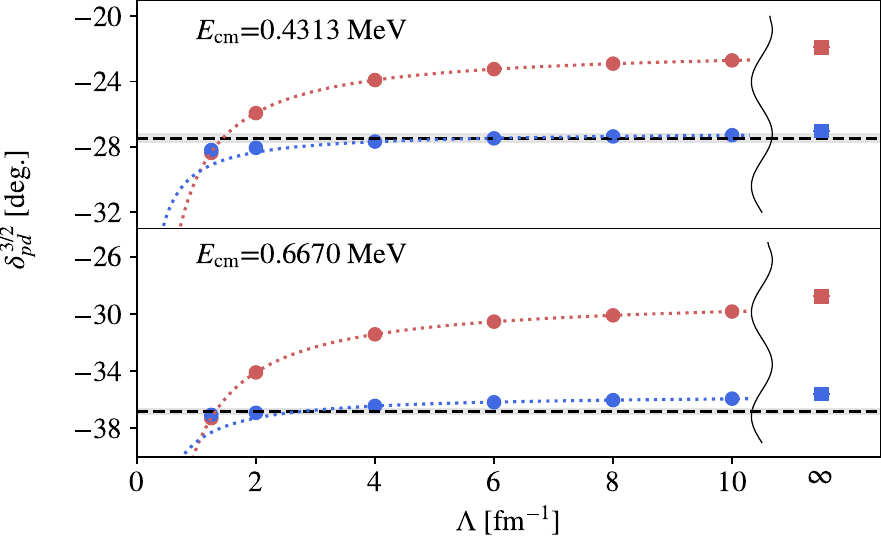}
    \end{tabular}
    \caption{The calculated $s$-wave $pd$ spin-quartet (left panel) and spin-doublet (right panel) scattering phase shifts, $\delta^{1/2}_{pd}$ and $\delta^{3/2}_{pd}$, respectively, as a function of the momentum cutoff $\Lambda$. The phase shifts are shown for two center-of-mass energies, $E_{cm}$, and compared to the most recent experimental PSA values \cite{PhysRevC.65.034002} denoted by the black dashed lines with the experimental error indicated by the gray band. The notation of our LO and NLO results is the same as in Fig.~\ref{pd_32_scatt_length}. For $\delta^{1/2}_{pd}$, the green circles show the results of our NLO calculations once the $pnn$ $D^{(1)}_0$ term is considered in place of the $ppn$ $D^{(1)}_1$ term.}
    \label{pd_scatter}
\end{figure*}

%-------------------------------------------------------------
\subsection{$\bm{p}\bm{d}$ scattering}
%-------------------------------------------------------------

As for \(pd\) scattering, there is a significant discrepancy among different experimental extractions of the \(pd\) spin-quartet scattering length. While the 1974 analysis~\cite{arvieux1974phase} reported a value of \(a^{3/2}_{pd} = 11.88^{+0.4}_{-0.1}\)~fm, and the 1983 analysis found \(a^{3/2}_{pd} = 11.11^{+0.25}_{-0.24}\)~fm, the most recent analysis from 1999~\cite{pd_scatt_length_black} suggests a substantially larger scattering length \(a^{3/2}_{pd} = 14.7(2.3)\)~fm, indicating a notable deviation from the earlier results. In theoretical studies, potential models predict values around $a^{3/2}_{pd}=13.6$~fm \cite{kievsky2008hh_review,hyperspherical_review,PhysRevC.44.50} in agreement with the most recent experimental datum. Within \nopieft the \(pd\) spin-quartet scattering length was calculated up to N2LO by K\"{o}nig et al. \cite{konig_pd}. They predict $a^{3/2}_{pd}=10.9(4)$~fm but argue that the discrepancy with potential models arises from different subtraction schemes for Coulomb effects. 

Our LO and NLO $\slashed{\pi}$EFT results for $a^{3/2}_{pd}$ are depicted in Fig.~\ref{pd_32_scatt_length} as functions of the cutoff. They were extracted from calculated phase shifts at $E_\text{cm}$ between 0.1 and 1~MeV. According to the adapted $\slashed{\pi}$EFT power counting, no three-body force is needed in the spin-quartet channel up to NLO. The same is true for the two-body forces projected to the spin-singlet channels. Consequently, none of the contact terms stemming from the inclusion of the Coulomb potential contribute to the spin-quartet case. We obtain
\begin{align}
    \text{LO:}\,\,&a^{3/2}_{pd}=9.9(1.1) \,\text{fm}, \quad r^{3/2}_{pd}=0.35(34) \,\text{fm}, \nonumber\\
    \text{NLO:}\,\,&a^{3/2}_{pd}=12.76(26) \,\text{fm}, \quad r^{3/2}_{pd}=1.16(8) \,\text{fm},
    \nonumber
\end{align}
where the uncertainty arises dominantly from the truncation error (errors, which stem from the few-body method or from fitting the effective range expansion to the \(pd\) phase shifts are negligible). The NLO result aligns with the most recent experimental datum from 1999~\cite {pd_scatt_length_black} and is consistent with the predictions of potential models.

Nucleon-deuteron $s$-wave scattering in the spin-doublet channel is characterized by having an 'anomalous' effective range expansion. In general, the amplitude may vanish for some complex momentum $k$ at some point other than the origin. Consequently, the phase shift is zero at this point and the convergence radius of the $k\cot{\delta}$ Taylor series is limited by the absolute value $k_0=|k|$. In spin-doublet $Nd$ scattering, $k_0$ happens to be close to zero, so that the usual effective range expansion used in Eqs.~\eqref{eq:nocoulomb_ere}~and~\eqref{eq:coulomb_ere} is no longer a useful parametrization of the scattering data. The expansion 
\begin{equation}\label{eq:ere_resonance}
    \kappa(\eta)=\frac{-1/a+1/2rk^2}{1\pm k^2/k_0^2},
\end{equation}
needs to be used instead \cite{eyre1977proton,chen1989low}. The sign in the denominator is plus if the pole is in the positive momentum plane and minus otherwise.

The $nd$ scattering length was already studied up to NLO $\slashed{\pi}$EFT in Ref.~\cite{schafer2023few}, and the authors found the predictions to be in good agreement with the experimental value and other theoretical predictions which hover around $a^{1/2}_{nd}=0.65$ fm. However, the $pd$ doublet scattering length is even harder to determine since the pole position is much closer to the $pd$ threshold, and highly accurate calculations (or measurements) for energies less than 0.1 MeV are needed to pinpoint its precise position \cite{pd_scatt_length_black}. Currently, experimental $a^{1/2}_{pd}$ extractions are not accurate \cite{pd_scatt_length_black} and theoretical results for various potentials contradict each other \cite{marcucci2020hyperspherical,kievsky2008hh_review}. It is not even clear whether the scattering length is positive or negative, but its value should be very close to zero. We made attempts to fit the modified ERE \eqref{eq:ere_resonance} to our phase shift calculations in the energy region $E_\text{cm}\in[0.1,1]$ MeV, but we were not able to pinpoint the pole position precisely to obtain values of the ERE parameters where large systematic errors would not weigh. For these reasons, we do not compare $a^{1/2}_{pd}$ to other calculations and experiments. Rather, we restrict our comparison with experimental data only to phase shifts at two selected energies below the deuteron breakup threshold.

Our results for the spin-quartet and spin-doublet $s$-wave phase shifts as functions of the cutoff $\Lambda$ are shown in Fig.~\ref{pd_scatter}. Here, calculations for two different center-of-mass energies $E_\text{cm}=0.4313$~MeV and $E_\text{cm}=0.6670$~MeV are presented and compared to values from the phase shift analysis (PSA) \cite{PhysRevC.65.034002}. At NLO, the agreement with experimental data for both energies is excellent in the $S=3/2$ channel, with a theoretical error estimated from the residual cutoff dependence being only about 1.5 degrees. Compared to the spin-quartet, the calculations for the spin-doublet are a bit more involved, since now pairs of nucleons couple to the total two-body spin either $S=1$ or $S=0$, which means that all two-body terms contribute. Furthermore, the LO three-body force and the NLO $ppn$ are required. Consequently, we also test if the NLO isospin-symmetry-breaking three-body force correctly renormalizes the calculated phase shifts. The calculations without this interaction are represented by the green points in the right panel of Fig.~\ref{pd_scatter}. Similarly to our study of the binding energy of $^3$He, shown in the left panel of Fig.~\ref{binding_energs}, the value of the phase shifts at the studied energies rises with the cutoff up to $\Lambda=12$ fm$^{-1}$, then steadily declines and does not converge. In contrast, the LO and NLO calculations, which include all forces mandated by the power counting, show the expected $\mathcal{O}(1/\Lambda)$ dependence on the residual cutoff. Interestingly, compared to the spin-quartet calculations, the spin-doublet phase shifts at NLO are more dependent on the cutoff, with a theoretical error of about 3 degrees for both energies. This indicates that $pd$ $S=1/2$ $s$-wave phase shifts converge rather slowly within the EFT expansion. Similar behavior was emphasized by Vanasse \emph{et al.} in Ref.~\cite{vanasse2014he}, where analytical calculations of $pd$ scattering observables were performed. The authors concluded that the inclusion of the NLO $ppn$ force leads to a larger cutoff dependence in this channel. Our results for the calculated phase shifts are listed in Tab.~\ref{tab:pd_phsf}.

\begin{table}
\begin{tabular}{lcccc}
\hline\hline
&~~~$E_\text{cm}$ [MeV]~~&LO&NLO&Exp.\\ \hline
$\delta^{1/2}_{pd}$&0.4313&-3.45(26)$^\circ$&-6.5(1.7)$^\circ$&-7.64(74)$^\circ$\\
&0.6670&-6.25(56)$^\circ$&-10.8(2.3)$^\circ$&-10.50(59)$^\circ$\\
$\delta^{3/2}_{pd}$& 0.4313&-21.9(2.0)$^\circ$&-27.24(44)$^\circ$&-27.48(25)$^\circ$\\
&0.6670&-28.8(2.7)$^\circ$&-35.87(56)$^\circ$&-36.81(19)$^\circ$\\
\hline
\hline
\end{tabular}
\caption{The calculated LO and NLO $pd$ $s$-wave phase shifts in the spin-doublet, $\delta^{1/2}_{pd}$, and spin-quartet, $\delta^{3/2}_{pd}$, channel, in the $\Lambda \rightarrow \infty$ contact limit, at two different center-of-mass energies $E_\text{cm}$. The corresponding total errors of LO and NLO results are given in the parentheses (for details, see the text). The listed experimental values are the most recent phase-shift analysis \cite{PhysRevC.65.034002}.}
\label{tab:pd_phsf}
\end{table}

We note that the larger error bars of spin-double phase shift at $\Lambda=18$ fm$^{-1}$ and {$\Lambda=~20$~fm$^{-1}$} stem from the uncertainty given by a very slow convergence of HO trap energies with the increasing amount of correlated Gaussian basis states. For such large cutoffs, we observe that it becomes increasingly challenging to obtain highly converged NLO perturbative corrections to our LO HO trap energies.

%-------------------------------------------------------------
\subsection{$S=2$ $\bm{d}\bm{d}$ scattering}
%-------------------------------------------------------------

There is only limited experimental data concerning the low-energy deuteron-deuteron scattering below the deuteron breakup threshold. The elastic $dd$ scattering was measured in Ref.~\cite{marlinghaus1975}, and the corresponding experimental cross-sections were subsequently analyzed using the resonating group method (RGM) employing a phenomenological $NN$ potential \cite{meier1975}.  In this theoretical study, the authors find that their calculations are in agreement with the experiment. By using the $s$-wave spin-quintet $dd$ phase shifts listed in Ref.~\cite{meier1975}, we deduce $a^2_{dd}=6.14\,\text{fm}$ and $r^2_{dd}=1.62\,\text{fm}$. From the purely theoretical side, $dd$ $s$-wave scattering in the spin-quintet channel was investigated using the Faddeev-Yakubovsky formalism in Refs.~\cite{fadeev_carew_2021}~and~\cite{filikhin2000}. In the first reference, a simple Gaussian potential is employed, and the reported value of the scattering length is $a^2_{dd}=7.8(3)$~fm. In the second reference, a Yukawa-type potential is used, specifically the so-called MT-I/III phenomenological model \cite{malfliet1969solution}. The reported scattering length is $a^2_{dd}=7.5$~fm.  

\begin{figure}[t!]
    \includegraphics[width=\columnwidth]{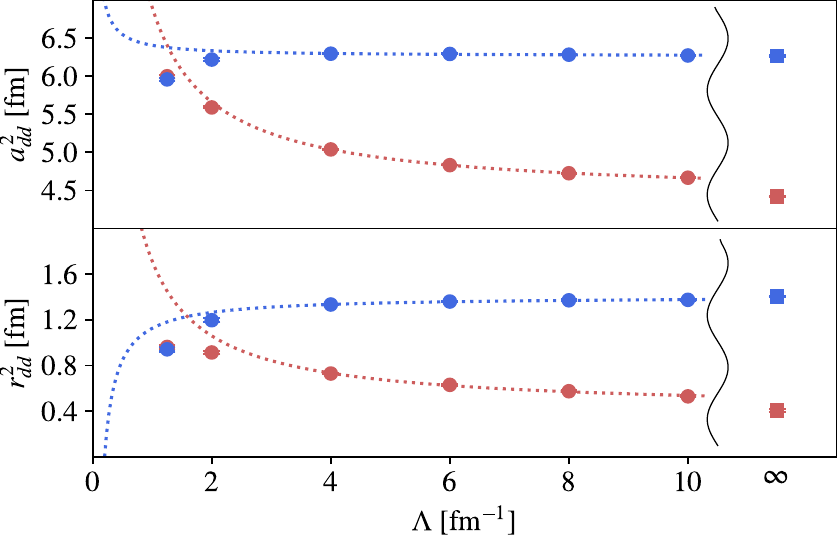}
    \caption{The calculated LO and NLO scattering length, $a^2_{dd}$, and effective range, $r^2_{dd}$,of $dd$ in the $S=2$ channel as a function of the momentum cutoff $\Lambda$. The notation is the same as in Fig.~\ref{pd_32_scatt_length}. }\label{fig:dd_s2_ere}
    \vspace{-10pt}
\end{figure}

Our results for the $dd$ scattering length and effective range in the $S=2$ channel at LO and NLO of $\slashed{\pi}$EFT are shown in Fig.~\ref{fig:dd_s2_ere}. Here, all nucleon pairs must couple to $S=1$, consequently, only the terms with $C_1^{(0)}$, $C_1^{(1)}$ and $C_4^{(1)}$ LECs contribute in this calculation. In this regard, the situation is somewhat similar to the $pd$ spin-quartet scattering. Our results for the ERE parameters  
\begin{align}
    &\text{LO}~~~:~a^2_{dd}=4.42(62)\,\text{fm},\quad r^2_{dd}=0.41(32)\,\text{fm},\nonumber\\
    &\text{NLO}~:~a^2_{dd}=6.262(42)\,\text{fm},\quad r^2_{dd}=1.41(7)\,\text{fm},\nonumber
\end{align}
slightly deviate from the Faddeev-Yakubovsky studies, but they agree with the earlier analysis using RGM.

%-------------------------------------------------------------
\subsection{$\bm{p}\bm{^3}$\textbf{He} scattering}
%-------------------------------------------------------------

\begin{figure*}
    \begin{tabular}[c]{cc}
            \includegraphics[width=\columnwidth]{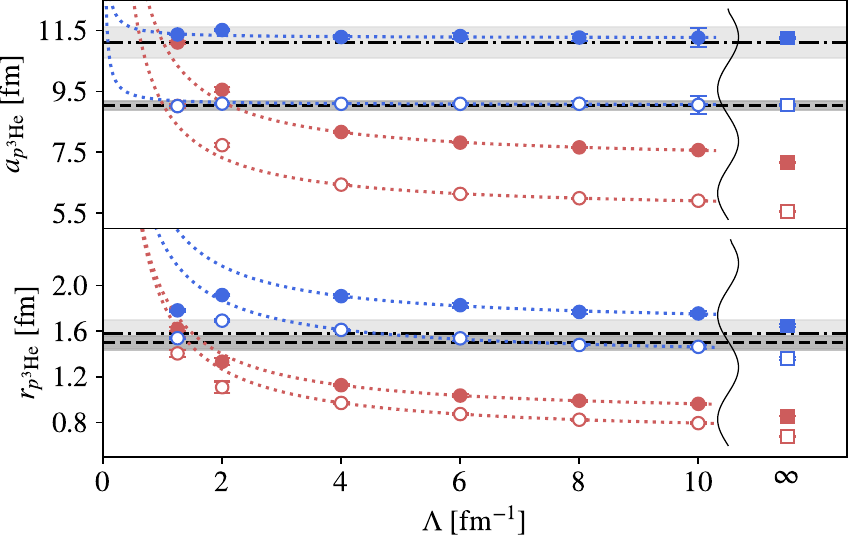}&
            \includegraphics[width=\columnwidth]{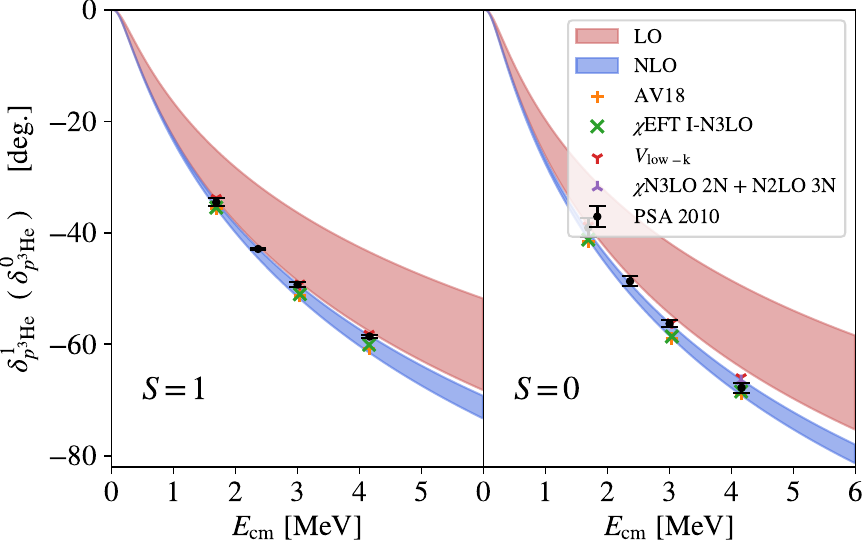}
    \end{tabular}
    \caption{The calculated LO and NLO spin-singlet ($S=0$) and spin-triplet ($S=1$) $p^3$He scattering lengths and effective ranges as a function of the increasing momentum cutoff $\Lambda$. The $S=0$ values are denoted by the filled circles and the $S=1$ values by the empty circles; apart from that, the notation is the same as in Fig.~\ref{pd_32_scatt_length}. The dashed-dotted and dashed black lines denote the corresponding spin-singlet and spin-triplet experimental values \cite{daniels2010spin}, respectively. The right panel: The spin-singlet $(S=0)$ and spin-triplet $(S=1)$ $p^3$He $s$-wave phase shifts evaluated for different center-of-mass energies $E_{\rm cm}$ using extracted LO and NLO values of the scattering lengths and effective ranges. LO values and their theoretical as well as numerical errors are given by the red band, and the NLO values are given by the blue band. The black error bars are the experimental data of the PSA \cite{daniels2010spin}. The results for the I-N3LO, AV18, $V_\text{low-k}$ and $\chi$EFT N3LO($2N$)+N2LO($3N$) potential models are from Refs. \cite{viviani2020,viviani2011benchmark}.}
    \label{p3He_scatter}
\end{figure*}

Much like the three-body $pd$ scattering, the four-body problem of scattering protons off $^3$He has presented a challenge for nuclear forces. It serves as an important benchmark, especially for the three-body nuclear interaction. Throughout the last few decades, various methods in combination with phenomenological as well as $\chi$EFT potentials have been used in its treatment. Notably, calculations using the Alt-Grassberger-Sandhas (AGS) equations with screened and renormalized Coulomb potential and various realistic $2N$ interactions were performed in Refs. \cite{deltuva2007four,deltuva2013calculation}, the results were found to be potential-model dependent. Furthermore, hyperspherical harmonics with the Kohn variational principle and $2N$ as well as $3N$ interactions were applied to this problem in Refs. \cite{hyperspherical_review, viviani2020, kievsky2008hh_review, viviani2013effect}. In the references \cite{viviani2020, viviani2013effect}, the effect of N2LO $\chi$EFT and AV18-IL7 $3N$ forces on $p^3$He scattering observables was investigated. Next, Faddeev-Yakubovsky calculations can be found in Ref. \cite{viviani2011benchmark}, where the three aforementioned methods were benchmarked with various nuclear potentials. 

Zero-energy $p^3$He scattering was studied using the correlated hyperspherical harmonics approach and the Kohn variational principle, with the $2N$ AV18 and $2N+3N$ AV18 + Urbana~IX potentials~\cite{viviani1998neutron}. Some older calculations of scattering lengths are summarized in Ref.~\cite{carbonell_review_2001}. Last but not least, $p^3$He scattering was also studied at LO $\slashed{\pi}$EFT using the RGM Ref.~\cite{kirscher2013zero}.

Together with the apparent interest of the theoretical community in $p^3$He scattering, there is also a large set of experimental data available (see e.g. the references in \cite{viviani2020}). Naturally, the scattering lengths and effective range determined from the experimental data are of particular interest to us. As of right now, the most recent PSA was performed by Daniels et al. in Ref.~\cite{daniels2010spin}. They obtained the ERE parameters $a^0_{p^3\text{He}}=11.1(5)\,\text{fm}$, $r^0_{p^3\text{He}}=1.58(12)\,\text{fm}$ in the spin-singlet channel and $a^1_{p^3\text{He}}=9.04(14)\,\text{fm}$, $r^1_{p^3\text{He}}=1.50(6) \,\text{fm}$ in the spin-triplet channel. Older analyses are available in Refs. \cite{tegner1983rate,alley1993effective,george2003scattering}, but due to a lack of experimental data at that time, there are large discrepancies in the obtained scattering length values.

Our $\slashed{\pi}$EFT results for the $p{}^3$He ERE parameters in the spin-singlet and spin-triplet channels are shown in the left panel of Fig.~\ref{p3He_scatter} as functions of the increasing momentum cutoff $\Lambda$. This four-body system is a stringent test of our interaction since, unlike in the case of $dd$ scattering, all terms of the potential except the four-body force $E_0^{(1)}$ and the three-body $pnn$ force $D_0^{(1)}$ play a role in the calculation. At NLO, we observe very little cutoff dependence in the two scattering lengths. Especially interesting is the residual cutoff dependence of $a^1_{p^3\text{He}}$, which is almost completely flat. This suggests that $a^1_{p^3\text{He}}$ is well-converged already at the next-to-leading order of the $\slashed{\pi}$EFT expansion. This is further supported by the fact that values for each cutoff sit within the grey band denoting the experimental datum and its error. The residual cutoff dependence of the effective ranges is larger, however, still in agreement with experimental values of PSA \cite{daniels2010spin}. 

The extrapolated results for the ERE parameters at LO read 
\begin{align}
    &a^0_{p^3\text{He}}=7.17(1.0)\,\text{fm},\quad r^0_{p^3\text{He}}=0.86(27)\,\text{fm}, \nonumber \\
    &a^1_{p^3\text{He}}=5.55(89)\,\text{fm},\quad r^1_{p^3\text{He}}=0.68(29)\,\text{fm}, \nonumber
\end{align}
where both values of scattering lengths agree well with the LO $\slashed{\pi}$EFT findings calculation of Ref. \cite{kirscher2013zero}. The NLO predictions are 
\begin{align}\label{eq:p3He_ere_results}
    &a^0_{p^3\text{He}}=11.26(4)\,\text{fm},\quad r^0_{p^3\text{He}}=1.65(26)\,\text{fm}, \nonumber\\
    &a^1_{p^3\text{He}}=9.06(4)\,\text{fm},\quad r^1_{p^3\text{He}}=1.36(25)\,\text{fm}. \nonumber
\end{align}

We insert the calculated values of $a_{p^3\text{He}}$ and $r_{p^3\text{He}}$ into the ERE \eqref{eq:coulomb_ere} and show the corresponding phase shifts for the spin-singlet and spin-triplet channels in the right panel of Fig.~\ref{p3He_scatter}. The red and blue bands are constructed by considering all possible cutoff values of phase shifts given by the running scattering lengths and effective ranges. The calculated phase shifts are compared to the PSA of Ref.~\cite{daniels2010spin} and other theoretical calculations. Taking into account the theoretical uncertainty, our calculations are in excellent agreement with the experimental phase shifts. 

In the spin-singlet channel, for energies below 3~MeV, only the phase shifts calculated with the non-local two-body $V_\text{low-k}$ interaction \cite{viviani2011benchmark}, agree with the PSA and our NLO results. The discrepancies of $\slashed{\pi}$EFT compared to the predictions of the hard-core AV18 and I-N3LO potentials~\cite{viviani2011benchmark} are about $1^\circ$ at 1.7 MeV. At higher displayed energies, all the compared phase shifts are within the NLO uncertainty band. In the spin-triplet channel, the calculated phase shifts stemming from different interaction models seem to be more consistent. The predictions of the AV18 and I-N3LO potentials are outside of our NLO uncertainty band only for the lowest energy phase shift. This discrepancy is about $0.7^\circ$.

%===================
\section{Conclusion}\label{sec:conclusion}
%===================
We presented a detailed analysis of charged bound systems and scattering processes involving up to four nucleons within the framework of \nopieft at NLO. The scattering problem is addressed by confining the interacting nuclei in a harmonic oscillator (HO) potential, which enables the extraction of phase shifts via a quantization condition, a generalized BERW formula that incorporates the long-range Coulomb interaction.

Our calculations are fully consistent with the power counting of \nopieft and exhibit renormalization group (RG) invariance. This is achieved by treating subleading contributions in first-order perturbation theory and by varying the momentum cutoff well beyond the breakdown scale of the theory given by the pion mass.

The \nopieft nuclear potential up to NLO, devised in this work, incorporates the Coulomb potential as a leading-order effect. This is consistent with the fact that at sufficiently low momenta the Coulomb interaction becomes non-perturbative and introduces a new dominant scale to the power-counting scheme. RG invariance is ensured by including the LO and NLO LECs that renormalize the $pp$ channel. We verified that a new isospin-symmetry-breaking three-body force needs to be included to renormalize the $s$-wave $S, I = 1/2,1/2$ $ppn$ channel properly at NLO. This result is consistent with the earlier findings of Ref.~\cite{vanasse2014he}. The primary advantage of our approach is that it can be easily extended to few- and many-body systems. 

We demonstrated the predictive capabilities of \nopieft by studying the $s$-wave $pd$, $p{}^3$He, and $dd$ scattering. The calculated NLO $pd$ spin-quartet scattering length agrees reasonably well with the predictions of various phenomenological models. Furthermore, it aligns with the most recent experimental extraction \cite{pd_scatt_length_black}, although the large experimental error associated with this datum makes it challenging to draw any conclusions.  At NLO, both $pd$ $s$-wave spin-quartet and spin-doublet phase shifts are in excellent agreement with the most recent PSA \cite{PhysRevC.65.034002}. 

We also studied the $s$-wave $dd$ $S=2$ effective range parameters. It is challenging to compare these predictions due to a lack of experimental data and calculations using realistic potentials in the spin-quintet channel. 

The NLO $s$-wave $p^3$He effective range parameters are in excellent agreement with the most recent experimental data \cite{daniels2010spin}. In particular,  the weak cutoff dependence of the $p^3$He scattering lengths suggests that they converge rather quickly with the increasing EFT order, and already at NLO, the residual cutoff dependence is almost flat. The $s$-wave $p^3$He phase shifts below the ${}^3$He breakup threshold are also in excellent agreement with the PSA~\cite{daniels2010spin} and either agree with or give predictions closer to the experimental values than other realistic potential models.  

Our calculations demonstrate that \nopieft, despite its apparent simplicity, is a theory with significant predictive power for nuclear processes at very low momenta. Of particular interest for future studies is the application of the interaction devised in this work to reactions of astrophysical interest, e.g., $dd$ fusion {\cite{Konrad2022}}, for which one of the elastic channels was already investigated in this work. The power-counting renormalizability of the \nopieft potential gives us a way to gauge theoretical errors and, as such, complements the experimental measurements, which are difficult to perform due to the strong low-energy Coulomb repulsion in these systems.

%========================
\section*{ACKNOWLEDGMENT}
%========================
The work of M.R. and M.S. was supported by the Czech Science Foundation GA\v{C}R grant 22-14497S. The work of M.B. and N. B. was supported by the Israel Science Foundation grant 1086/21.  

%=========
\appendix*
%=============
\section{Low Energy Constants}\label{sec:lecs}
%=============
The LECs used throughout this work are given in Tabs.~\ref{tab:lo_lecs_2b},~\ref{tab:nlo_lecs_2b}, and \ref{tab:nlo_lecs_3b_4b}. Here, we shall also describe the fitting procedure of the $pp$ two-body LECs in more detail. The fitting is done only for the two-body orbital momentum $l=0$ ($s$-wave); however, the description is given for a general partial wave. The full Coulomb plus short-range potential partial-wave amplitude is given by 
\begin{equation}\label{eq:partial_wave_amp}
    f_l(k)=\frac{e^{2i\sigma_l}-1}{2ik} + e^{2i\sigma_l}\frac{e^{2i\delta_l}-1}{2ik}, 
\end{equation}
where $\sigma_l$ is the pure Coulomb phase shift defined by $\sigma_l=\text{arg}\,\Gamma(1+l+i\eta)$.
The first term of \eqref{eq:partial_wave_amp} is the pure Coulomb amplitude $f_l^{(c)}$, which is known exactly and is of no interest to us. The second term is the additional amplitude due to the nuclear interaction, and it is easy to show that 
\begin{equation}\label{eq:real_amp}
    \text{Re}\left[\frac{e^{2i\sigma_l}}{f'_l(k)}\right]=k\cot\delta_l,
\end{equation}
where $f'_l=f_l-f_l^{(c)}$.
At LO, we need to calculate the amplitude $f'_l$ or, equivalently, the $S$-matrix element. It is given by the ratio of an outgoing and incoming Jost function 
\begin{equation}
    S_l(k)=e^{2i(\delta_l+\sigma_l)}=\frac{\mathcal{F}_l(-k)}{\mathcal{F}_l(+k)}e^{i\pi l}, 
\end{equation}
where the Jost function $\mathcal{F}_l$ is defined as 
\begin{equation}\label{eq:jost_integral}
    \mathcal{F}_l(k)=\mathcal{F}^{(c)}_l(k)+2\mu \int_{0}^{\infty}dr\,u_l^{(c)}V^{(0)}(r)\varphi_l(r), 
\end{equation}
where $\mathcal{F}^{(c)}_l(k)$ is the "pure Coulomb Jost function" given by 
\begin{equation}
    \mathcal{F}^{(c)}_{l,\pm}(k)=\frac{(2k)^{-l}e^{\frac{1}{2}\eta\pi\pm i\frac{1}{2}\pi l}\Gamma(2l+2)}{\Gamma(l+1\pm i\eta)}
\end{equation}
and the function $u_l^{(c)}$ is defined as 
\begin{equation}
    u_{l,\pm}^{(c)}=e^{\frac{1}{2}\pi\eta}\mathcal{W}_{\mp i\eta, l+1/2}(\mp2ikr).
\end{equation}
The wavefunction $\varphi_l$ in \eqref{eq:jost_integral} is the regular solution of the two-body Schr\"{o}dinger equation with $V_c+V^{(0)}$. We then find such LO LECs so that the ERE at the right-hand side of \eqref{eq:real_amp} reproduces the desired scattering length. 

At NLO, we calculate the first-order perturbative correction $\delta f'_l$ to the amplitude $f'_l$. This is done using the distorted-wave Born approximation. We find that 
\begin{equation}
    \delta f'_l = -\frac{2\mu}{\mathcal{F}_l(k)^2}~e^{i\pi l}\int_0^\infty dr\,\varphi_l(r)V^{(1)}(r)\varphi_l(r). 
\end{equation}
Then, similarly to LO, the NLO LECs of the potential $V^{(1)}$ are adjusted so the amplitude $f'_l+\delta f'_l$ reproduces the desired effective range while keeping the scattering length the same. 

\begin{table*}
    \centering
    \caption{The fitted values of LO LECs for different momentum cutoffs $\Lambda$ used in this work. For more details, see Sec.~\ref{sec:theory}.}
    \label{tab:lo_lecs_2b}
    \begin{tabular}{crrrr}
        \hline
        $\Lambda$ [fm$^{-1}$]& $C^{(0)}_0$ [MeV]~~~~& $C^{(0)}_1$ [MeV]~~~~& $C^{(0)}_{2}$ [MeV]~~~~& $D^{(0)}_0$ [MeV]~~~~\\[2pt]
        \hline
        \hline 
    1.00	&-2.4865368E+01	&-4.4453335E+01	&-2.4503698E+01	&-1.0902644E+00\\
    1.25	&-3.9701746E+01	&-6.3713715E+01	&-3.9796207E+01	&7.0398708E+00\\
    2.00	&-1.0507232E+02	&-1.4236875E+02	&-1.0722713E+02	&6.5401277E+01\\
    3.00	&-2.4092326E+02	&-2.9594289E+02	&-2.4686441E+02	&2.5992292E+02\\
    4.00	&-4.3242573E+02	&-5.0517262E+02	&-4.4292988E+02	&6.6134375E+02\\
    5.00	&-6.7958136E+02	&-7.7005720E+02	&-6.9521141E+02	&1.3819823E+03\\
    6.00	&-9.8239072E+02	&-1.0905964E+03	&-1.0035898E+03	&2.5882872E+03\\
    7.00	&-1.3408540E+03	&-1.4667900E+03	&-1.3679877E+03	&4.5316047E+03\\
    8.00	&-1.7549714E+03	&-1.8986379E+03	&-1.7883512E+03	&7.5954620E+03\\
    9.00	&-2.2247429E+03	&-2.3861403E+03	&-2.2646405E+03	&1.2369085E+04\\
    10.00	&-2.7501685E+03	&-2.9292970E+03	&-2.7968247E+03	&1.9762682E+04\\
    12.00	&-3.9679825E+03	&-4.1825733E+03	&-4.0287842E+03	&4.8855177E+04\\
    14.00	&-5.4084134E+03	&-5.6584666E+03	&-5.4840820E+03	&1.1874743E+05\\
    16.00	&-7.0714614E+03	&-7.3569774E+03	&-7.1626172E+03	&2.8914474E+05\\
    18.00	&-8.9571262E+03	&-9.2781052E+03	&-9.0643066E+03	&7.1309616E+05\\
    20.00	&-1.1065408E+04	&-1.1421850E+04	&-1.1189090E+04	&1.7935213E+06\\
        \hline
    \end{tabular}
\end{table*}

\begin{table*}
    \centering
    \caption{The fitted values of two-body NLO LECs for different momentum cutoffs $\Lambda$ used in this work. For more details, see Sec.~\ref{sec:theory}.}
    \label{tab:nlo_lecs_2b}
    \begin{tabular}{ccccccc}
        \hline
        $\Lambda$ [fm$^{-1}$]& $C^{(1)}_0$ [MeV] & $C^{(1)}_1$ [MeV.fm$^{-2}$] & $C^{(1)}_2$ [MeV] & $C^{(1)}_3$ [MeV.fm$^{-2}$] & $C^{(1)}_4$ [MeV] & $C^{(1)}_5$ [MeV.fm$^{-2}$]\\[2pt]
        \hline
        \hline 
    1.00	&1.9136458E+00	&3.4599591E+00	&8.8632497E+00	&8.4357420E+00	&9.5218829E-01	&1.9704547E+00\\
    1.25	&-3.7041922E+00	&-4.1706419E+00	&1.5330531E+00	&1.0111918E+00	&-5.2853807E+00	&-6.4866069E+00\\
    2.00	&-6.7162510E+01	&-2.8316785E+01	&-6.5124557E+01	&-1.9200349E+01	&-7.6729955E+01	&-3.3224738E+01\\
    3.00	&-3.3654713E+02	&-6.1559657E+01	&-3.0501565E+02	&-4.3506934E+01	&-3.7816882E+02	&-6.9373331E+01\\
    4.00	&-9.3642545E+02	&-9.5175229E+01	&-7.9345153E+02	&-6.6638744E+01	&-1.0417942E+03	&-1.0530159E+02\\
    5.00	&-1.9973117E+03	&-1.2896064E+02	&-1.6138938E+03	&-8.9265132E+01	&-2.2065976E+03	&-1.4118151E+02\\
    6.00	&-3.6495287E+03	&-1.6282773E+02	&-2.8499217E+03	&-1.1163142E+02	&-4.0039381E+03	&-1.7672942E+02\\
    7.00	&-6.0234489E+03	&-1.9674193E+02	&-4.5851446E+03	&-1.3384690E+02	&-6.5736839E+03	&-2.1226093E+02\\
    8.00	&-9.2494482E+03	&-2.3068582E+02	&-6.9031826E+03	&-1.5596731E+02	&-1.0049478E+04	&-2.4770660E+02\\
    9.00	&-1.3457901E+04	&-2.6464961E+02	&-9.8876604E+03	&-1.7802400E+02	&-1.4566198E+04	&-2.8308435E+02\\
    10.00	&-1.8779180E+04	&-2.9862736E+02	&-1.3622205E+04	&-2.0003593E+02	&-2.0258537E+04	&-3.1840716E+02\\
    12.00	&-3.3281723E+04	&-3.6661092E+02	&-2.3676004E+04	&-2.4396996E+02	&-3.5708589E+04	&-3.8892910E+02\\
    14.00	&-5.3800076E+04	&-4.3461863E+02	&-3.7733627E+04	&-2.8782694E+02	&-5.7475272E+04	&-4.5932611E+02\\
    16.00	&-8.1377243E+04	&-5.0264150E+02	&-5.6464054E+04	&-3.3163529E+02	&-8.6627957E+04	&-5.2959978E+02\\
    18.00	&-1.1705705E+05	&-5.7067855E+02	&-8.0536374E+04	&-3.7541151E+02	&-1.2424618E+05	&-5.9982045E+02\\
    20.00	&-1.6189814E+05	&-6.3878611E+02	&-1.1061960E+05	&-4.1916510E+02	&-1.7140531E+05	&-6.7000446E+02\\
        \hline
    \end{tabular}
\end{table*}

\begin{table*}
    \centering
    \caption{The fitted values of three- and four-body NLO LECs for different momentum cutoffs $\Lambda$ used in this work. For more details, see Sec.~\ref{sec:theory}.}
    \label{tab:nlo_lecs_3b_4b}
    \begin{tabular}{crrr}
        \hline
        $\Lambda$ [fm$^{-1}$]~~~~& $D^{(1)}_0$ [MeV]~~~~& $D^{(1)}_1$ [MeV]~~~~& $E^{(1)}_0$  [MeV]~~~~\\[2pt]
        \hline
        \hline 
    1.00	&4.5427770E+00	&3.7345921E+00	&9.9330531E+01\\
    1.25	&-2.9117896E+00	&-4.2657522E+00	&1.7297277E+02\\
    2.00	&-9.7494030E+00	&-1.1865454E+01	&5.5493105E+03\\
    3.00	&2.9945647E+02	&3.0986944E+02	&1.6175396E+05\\
    4.00	&1.8243328E+03	&1.8920487E+03	&1.3212171E+06\\
    5.00	&6.3182933E+03	&6.5451499E+03	&3.9641260E+06\\
    6.00	&1.6955221E+04	&1.7533782E+04	&-1.6151691E+07\\
    7.00	&3.9275626E+04	&4.0566850E+04	&-2.9926015E+08\\
    8.00	&8.2770879E+04	&8.5410766E+04	&-2.4890734E+09\\
    9.00	&1.6353276E+05	&1.6862362E+05	&-1.6190575E+10\\
    10.00	&3.0869683E+05	&3.1812520E+05	&-9.2502866E+10\\
    12.00	&1.0059802E+06	&1.0360329E+06	&-1.8141796E+12\\
    14.00	&3.0571786E+06	&3.1473390E+06	&-3.7140906E+12\\
    16.00	&8.9868727E+06	&9.2505922E+06	&-5.2331519E+12\\
    18.00	&2.6126250E+07	&2.6893632E+07	&-9.2088287E+12\\
    20.00	&7.6168337E+07	&7.8413315E+07	&-1.8024255E+13\\
        \hline
    \end{tabular}
\end{table*}

\FloatBarrier
%============================================================================

\bibliography{refs} % Entries are in the refs.bib file
\end{document}